\documentclass{sig-alternate-05-2015}

\usepackage{epsfig}

\usepackage{graphicx}   

\usepackage{cite}

\usepackage{booktabs}

\usepackage{balance}  
\usepackage{amsmath}
\usepackage{xspace}
\usepackage{subfigure}
\usepackage{url}
\usepackage{t1enc,dfadobe}

\usepackage{algorithm}
\usepackage{algpseudocode}
\usepackage{comment}

\newif\ifremark
\long\def\remark#1{
\ifremark%
	\begingroup%
	\dimen0=\columnwidth
	\advance\dimen0 by -0.25in%
	\setbox0=\hbox{\parbox[b]{\dimen0}{\protect\em #1}}
	\dimen1=\ht0\advance\dimen1 by 2pt%
	\dimen2=\dp0\advance\dimen2 by 2pt%
	\vskip 0.25pt%
	\hbox to \columnwidth{%
		\vrule height\dimen1 width 3pt depth\dimen2%
		\hss\copy0\hss%
		\vrule height\dimen1 width 3pt depth\dimen2%
	}%
	\endgroup%
\fi}

\remarktrue

\newcommand{\anywaredc}{AnywareDC}
\newcommand{\anywareups}{AnywareUPS}

\begin{document}\sloppy
 \topmargin=0mm

\title{A Green Enterprise Computing Architecture for Developing Countries}

\numberofauthors{2}

\author{
%
%
\alignauthor
Rabia Akbar\\
       \affaddr{National University of Sciences and Technology (NUST)}\\
       \affaddr{School of Electrical Engineering and Computer Science (SEECS)}\\
       \affaddr{Islamabad, Pakistan}\\
       \email{12mscsrakbar@seecs.edu.pk}
\alignauthor
       Tahir Azim\\
       \affaddr{National University of Sciences and Technology (NUST)}\\
       \affaddr{School of Electrical Engineering and Computer Science (SEECS)}\\
       \affaddr{Islamabad, Pakistan}\\
       \email{tahir.azim@seecs.edu.pk}
}

\maketitle

\begin{abstract}

Developing countries often have access to limited energy resources, which 
frequently results in power cuts and failures. During these power cuts,
enterprises rely on backup sources for power such as uninterruptible power 
supplies (UPS) and electric generators. 
This paper proposes \anywaredc{}, 
an architecture that builds on the recent work on Anyware~\cite{kazandjieva2014system} 
to reduce energy utilization in the presence of such intermittent power supplies.
Anyware reduces energy usage by providing enterprise users laptops instead of desktops, while maintaining 
performance using a central compute cluster. Our basic insight is that in the presence of 
power cuts, only the routers and the cluster needs to be provided power; the laptops can continue 
to run on their own batteries. This reduces both energy usage and UPS load allowing it to supply power for longer, 
thus also saving generator fuel costs. Simulations show that this 
architecture reduces energy usage by up to 80\% compared to one not using Anyware, and by up to 
20\% compared to Anyware.

\end{abstract}




\section{Introduction}

Computing equipment is one of the major energy sinks in corporate and academic 
buildings today \cite{kazandjieva2012green}. While computers enable remarkable 
increases in productivity, their energy costs have continued to increase over 
the past several years. Developing countries can benefit greatly from this
higher productivity, but they often have limited energy resources. 
This makes it crucial to utilize the available resources as efficiently as 
possible. Furthermore, limited energy resources often also lead to rolling 
blackouts and sometimes prolonged power cuts, during which backup power sources 
such as uninterrupted power supplies (UPS) and electric generators are needed 
to keep the computing equipment running.

To increase the energy efficiency of computing equipment in an enterprise, 
the Anyware project \cite{kazandjieva2014system}
recently proposed that users should be provided only laptops at work. 
To enable users to run more 
CPU-intensive programs, Anyware connects the laptops to a server cluster that executes
such programs transparently from the user. 
Since one server can provide computational support to many laptops, the 
total energy utilization is reduced.  

This paper presents Anyware for Developing Countries (\anywaredc), an architecture that supports
energy-efficient computing in the presence of frequent and possibly prolonged power cuts. We propose that by connecting 
the UPS only to the server cluster and the networking equipment, we can avoid expending its stored energy 
on a whole host of computers spread throughout the office. The laptops can continue running on their own 
batteries without either UPS or main power supply. This could allow the UPS to 
continue running for a much longer period of time, thus increasing its battery 
life and reducing the need to switch to generators.

Our simulation results show that this architecture results in up to 80\% 
reduction in energy consumption compared to a traditional desktop-based system, and a 
20\% reduction over a system using the basic version of Anyware.

The next section provides a brief introduction to Anyware. Section 
\ref{sec:anywaredc} describes our proposed
architecture in more detail. Section \ref{sec:evaluation} presents a simulation-based 
evaluation, including potential energy savings and the 
conditions in which this architecture could be beneficial. Finally, we describe
related work and conclude.

\section{Anyware}
\label{sec:anyware}

Anyware~\cite{kazandjieva2014system} is a system architecture that reduces the energy consumption of 
computing infrastructure in office buildings. It achieves this without 
sacrificing performance or putting devices to sleep. Anyware leverages the fact 
that laptops consume much less energy than desktop machines. 
Therefore, it suggests that every user in an office building should be provided 
a laptop for work purposes instead of a desktop. 

However, laptops also have much 
lower compute power than desktop machines. As a result, users prefer 
to use desktop machines in case they ever need to run 
compute-intensive tasks. 
Anyware solves this problem by proposing a centralized server cluster hosting 
virtual machines for users. Each user has his or her own virtual machine in the 
cluster. Each virtual machine has a software configuration that is identical to 
the laptop provided to the user. When a user working on a laptop starts
a task, an Anyware daemon running on the laptop intercepts the user request 
and decides if the task would run faster on the laptop itself or on the 
cluster. In general, compute-intensive tasks end up running transparently on the 
cluster, giving users the illusion of having powerful computational 
resources at their finger tips.

To achieve reasonable performance, the server cluster has to  
reside in the same network as the laptops. Each server is capable of 
hosting virtual machines for up to 25 users, thus 
making the server cluster much smaller than the number of laptops. 
User studies show that the process of making the 
decision to run locally or remotely, and then actually executing the task is 
completely transparent: users are unable to distinguish between a process 
running locally and one running remotely. Experimental evaluations further
show that this architecture allows enterprise computing equipment to run at 
20\% the cost of a desktop machine-based architecture.

\section{\anywaredc{}}
\label{sec:anywaredc}

In this paper, we propose \anywaredc{}, an architecture that extends Anyware to 
deal with environments where power outages are common. The existing Anyware 
architecture is designed for environments where there is a constant, 
uninterrupted supply of power. As a result, it makes no extra attempt to 
conserve energy during periods of power loss.

A naive way of extending Anyware to deal with power interruptions is to simply
connect the entire IT infrastructure to a UPS and generator system. We call this
simple approach \anywareups{}. The UPS 
would keep everything running when the power goes out, while the generator 
provides backup power if the power outage is prolonged and the UPS' batteries
start getting depleted. While this approach would work, it does not improve
energy efficiency compared to Anyware. The extra power needed to charge the UPS
batteries adds overhead to the entire system.

\anywaredc{} leverages the fact that laptops have their own built-in power 
supplies, so they do not need any backup power supplies in case of a power outage.
Instead, the backup power source provided by the UPS should be used only 
to power Anyware's server cluster and networking equipment such as routers and 
switches. If the UPS battery or one of the laptops' batteries start running out, 
only then does the generator need to be powered up.

\begin{figure}[h!]
\centering
\includegraphics[width=0.5\textwidth]{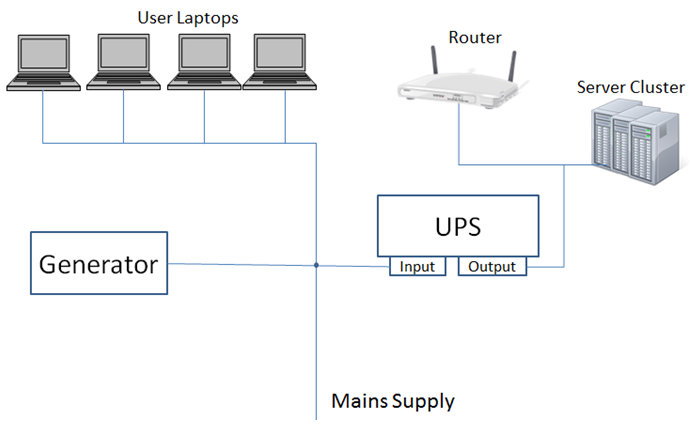}
\caption{Architectural diagram for \anywaredc{}.}
\label{fig:anyware-arch}
\end{figure}

Figure \ref{fig:anyware-arch} shows a simple architecture implementing this idea. The mains 
electrical supply is provided as input to the UPS as well as to the laptops 
provided to each user. However, the UPS output is supplied only to the central 
server cluster and the networking equipment, because the laptops can continue 
running on their own batteries. The generator acts as a backup power supply when 
the UPS or one of the laptops starts running out of power. A simple power 
monitoring tool running on the UPS and each of the users' laptops can be used to 
automatically trigger generator startup. If some laptops have much
lower battery lives than others, we suggest external battery packs to keep them
running or simply investing in newer laptops.

Since the server cluster is small and there are much fewer routers and switches
than laptops, the energy consumption can be greatly reduced. 
This can allow the UPS to continue supplying power for a much longer period of time. 
Moreover, longer UPS operation would lead to less frequent utilization of the 
generator, thus saving generator fuel costs. Sections
\ref{sec:energy_savings} and \ref{sec:generator_fuel_savings} validate these hypotheses. 

Of course, this architecture is not beneficial if the laptop batteries are weak and last only a few 
minutes. This would cause the generator to start up earlier than it would 
without this architecture. Section \ref{sec:laptop_battery_effect} evaluates the
minimum battery power laptops need to make this architecture beneficial.

\section{Evaluation}
\label{sec:evaluation}

This section evaluates the benefits and preconditions of using \anywaredc{}
in the presence of an intermittent power supply.
We first demonstrate the increase in
energy efficiency achieved by using \anywaredc{} instead of using only Anyware or
\anywareups{}.
Next, we show savings in generator fuel achieved using \anywaredc{}.
Finally, we evaluate the effect of laptop and UPS 
batteries on the overall effectiveness of \anywaredc{}.

The evaluation uses a discrete event simulator which models time in 
units of hours. Every hour, the simulator computes the energy consumed by 
various components of the computing infrastructure. Based on a pre-specified
policy, it simulates interruptions in power supply in any given hour 
and computes changes in energy consumption accordingly. The simulation is based on 
a small office environment with 25 users, with typical power draws for various
kinds of computing equipment at 100\% CPU usage: 24W per laptop, 
165W per desktop, 270W per server, 6W per switch, and 96W for the UPS. 
The generator is assumed to be powerful enough to be able to run the
entire computing infrastructure. 

\subsection{Energy Savings using \anywaredc{}}
\label{sec:energy_savings}

We first show the benefit of using \anywaredc{} compared to using
Anyware unchanged in an office computing environment. We use a random power
outage policy where, at the top of every hour, a power outage is simulated
based on some probability. The simulation assumes a typical scenario where
all laptops and the UPS have 3-hour battery backups.

\begin{figure}[h!]
\centering
\includegraphics[width=0.5\textwidth]{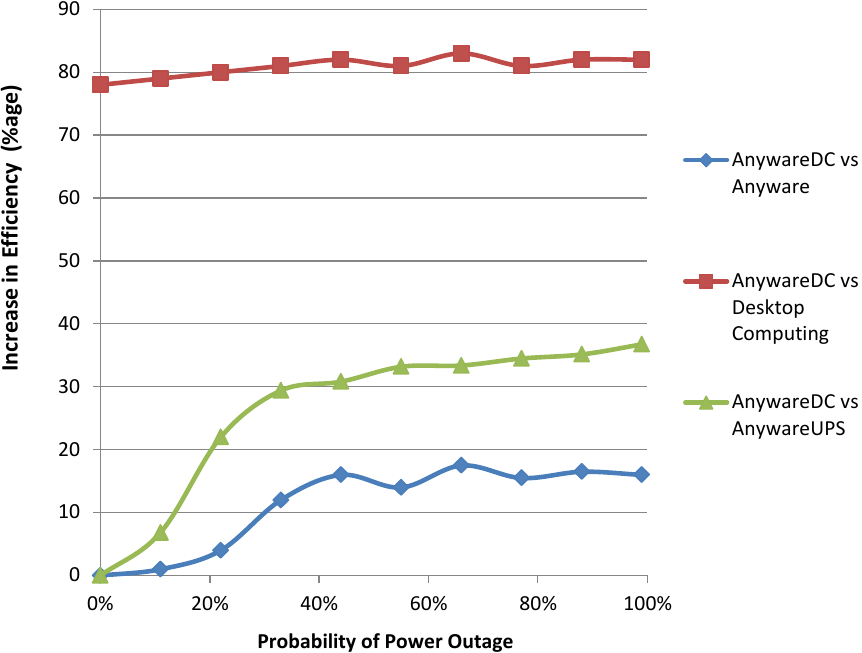}
\caption{Energy efficiency of \anywaredc{} compared to unmodified Anyware and a typical
desktop-dominated computing environment.}
\label{fig:anyware-efficiency}
\end{figure}

Figure \ref{fig:anyware-efficiency} plots the
energy efficiency achieved using \anywaredc{} for different power outage
probabilities. For probability below 20\%, there is little to no benefit 
because the duration where the laptops save energy by using their own batteries
is very small. For higher power outage probabilities, \anywaredc{} provides
between 15-20\% energy savings compared to unmodified Anyware. Furthermore,
similar to unmodified Anyware, \anywaredc{} achieves almost 80\% reduction in 
energy consumption compared to a typical desktop-dominated computing environment.

Of course, in order to make Anyware work in an intermittent power environment, 
we would have to install some backup power source to keep the infrastructure
running. Section \ref{sec:anywaredc} describes \anywareups{}, 
a simple method of achieving this using a UPS and generator system. However
the presence of the UPS adds extra energy overhead to the entire system. Figure
\ref{fig:anyware-efficiency} shows that due to this overhead of \anywareups{}, 
\anywaredc{} is able to improve on it by 30-35\% in terms of energy efficiency.

\subsection{Reduction in Generator Usage}
\label{sec:generator_fuel_savings}

\begin{figure}[h!]
\centering
\includegraphics[width=0.5\textwidth]{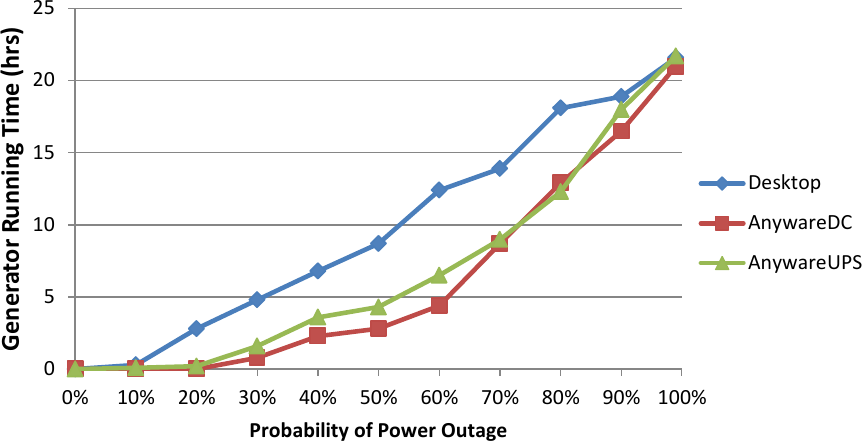}
\caption{Hours of generator use by \anywaredc{}, \anywareups{} and
a desktop-based enterprise for varying power outage probabilities.}
\label{fig:anyware-generator-fuel-usage}
\end{figure}

Secondly, we evaluate the reduction in generator usage achieved using
\anywaredc{}. The results are shown in Figure \ref{fig:anyware-generator-fuel-usage}.
In short, an architecture using \anywaredc{} requires almost 30\%
less generator fuel compared to \anywareups{} over a wide range of
power outage probabilities. This is 
because in \anywaredc{}, the UPS can backup the server cluster for a longer period of time. As a
result, the generator has to be run less frequently, resulting in
lower fuel consumption.

\subsection{Effect of Battery Backups}
\label{sec:laptop_battery_effect}

The effectiveness of \anywaredc{} heavily depends on the presence of 
reasonable battery backups on the UPS and laptop machines. If the UPS has weak
batteries that immediately get discharged, the infrastructure will have to switch
over to generator power immediately. Similarly, if a critically important laptop
loses battery power too quickly, either the generator will have to keep them running
or the laptop's user will have to wait until power is restored.

\begin{figure}[h!]
\centering
\includegraphics[width=0.5\textwidth]{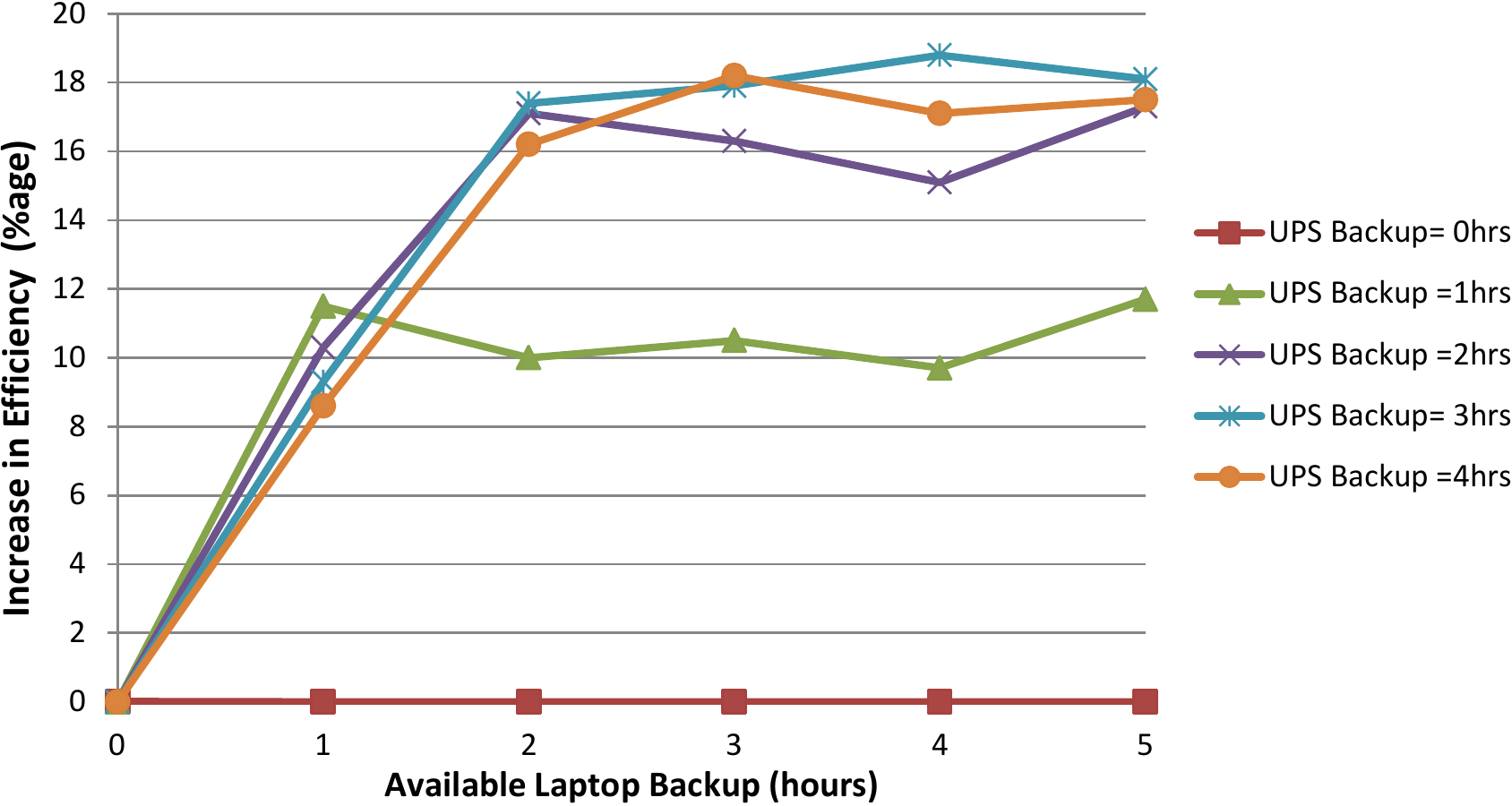}
\caption{Effect of laptop and UPS battery backup on the energy efficiency of
\anywaredc{}. Each hour, the probability of power availability is 50\%.}
\label{fig:anyware-ups-laptop-battery}
\end{figure}

This section explores the effect of laptop and UPS battery backups on the 
effectiveness of \anywaredc{}. Assuming a random power outage probability of 50\%,
Figure \ref{fig:anyware-ups-laptop-battery} illustrates the effect of laptop and 
UPS battery backups on the energy efficiency of \anywaredc{}. As we can see, if
there is no laptop battery backup, \anywaredc{} provides no real benefit
compared to the unmodified Anyware. The reason is that the generator needs to
be switched on immediately to provide power to the laptops. However, even a laptop
battery lasting 1 hour yields around 10\% improvement in energy efficiency. This
efficiency further increases to almost 20\% for longer battery lives.

The effect of UPS battery backup is similar. With no UPS batteries, the infrastructure
must switch to generator backup immediately. This leads to no improvement in energy efficiency. 
However, with a longer UPS battery life, the infrastructure can continue running
without generator backup for much longer, resulting in higher efficiency.

\section{Related Work}
\label{sec:related}

A substantial amount of work exists on reducing the energy consumption of computing
infrastructures. \cite{orgerie2014survey} presents an extensive survey of some of
the most common techniques. In summary, work on computing energy efficiency can be
classified into node-level optimizations, datacenter power management, 
and virtualization.

Node-level optimizations focus on optimizing the energy usage of a 
single PC or server. They often involve switching a PC to sleep mode when 
idle ~\cite{Reich2010, Sen2012-GreenUp}. They could
also involve hardware ~\cite{snowdon2005power} and software optimizations for 
a single node \cite{blanquicet2008managing}.

Datacenter power management techniques include setting up datacenters near 
green sources of electricity, such as hydel and wind power. More sophisticated
techniques run jobs on as few nodes as possible in a data center to reduce
energy usage \cite{chase2001balance}. Anyware uses a similar approach of running expensive jobs on
a very small server cluster.

Virtualization approaches, such as CloneCloud~\cite{clonecloud}, 
enable applications to be centralized and run in
application-layer VMs. Some approaches combine sleep-based techniques with virtualization.
SleepServer~\cite{Agarwal2010}, for example,
proxies applications in trimmed-down virtual machines, while 
LiteGreen~\cite{Das2009} runs the entire
user desktop environment in a virtual machine, which can be migrated
to a server so the desktop can be put to sleep.

\section{Conclusion and Discussion}
\label{sec:discussion}

This paper describes \anywaredc{}, a system architecture for higher energy efficiency in the
presence of intermittent power supply in developing countries. 
Besides its energy efficiency benefits, \anywaredc{}
also has other tangible benefits. First, it can simplify  administration of the IT
infrastructure. During power outages, since only a central cluster and a handful
of networking devices consume power, administrators only have to worry about their
power usage. This makes it easier to identify power hogs, detect malfunctions and failures,
and balance load across these devices. Secondly, laptop batteries last longer because they are regularly charged and
discharged during power outages. This keeps them in better running condition than if
they remain on main power supply all the time.
Finally, UPSes remain in better working condition since they drive a lighter
power load. This reduces the chances of batteries getting fully discharged or the UPS
getting damaged due to overload.

\bibliographystyle{abbrv}

\bibliography{ias}

\end{document}